\documentclass[sigchi, review=false]{acmart}
\usepackage{booktabs} % For formal tables

\usepackage[utf8]{inputenc}
\usepackage[table]{ colortbl}
\usepackage{xparse,xstring}
\usepackage{color,soul}
\usepackage{csquotes}
\usepackage{booktabs,rotating,makecell,multirow}
\usepackage{graphicx}
\usepackage{hyperref}
\usepackage[export]{adjustbox}
\usepackage{microtype}
\usepackage{flushend}

\definecolor{blue}{rgb}{0.88,0.88,1}

\begin{document}

\setlength{\intextsep}{5pt}
\setlength{\textfloatsep}{1pt}
\newcommand{\squeezeup}{\vspace{-2.5mm}}

\newcommand*{\belowrulesepcolor}[1]{%
  \noalign{%
    \kern-\belowrulesep
    \begingroup
      \color{blue}%
      \hrule height\belowrulesep
    \endgroup
  }%
}
\newcommand*{\aboverulesepcolor}[1]{%
  \noalign{%
    \begingroup
      \color{blue}%
      \hrule height\aboverulesep
    \endgroup
    \kern-\aboverulesep
  }%
}

\settopmatter{printacmref=true}
\title{The Heat is On: Exploring User Behaviour in a Multisensory Virtual Environment for Fire Evacuation}
%\titlenote

\author{Emily Shaw}
\affiliation{%
 \institution{University of Nottingham}
  \city{Nottingham, UK}
}
\email{Emily.Shaw1@nottingham.ac.uk}

\author{Tessa Roper}
\affiliation{%
 \institution{University of Nottingham}
   \city{Nottingham, UK}
}
\email{Tessa.Roper@nottingham.ac.uk}

\author{Tommy Nilsson}
\affiliation{%
 \institution{University of Nottingham}
  \city{Nottingham, UK}
}
\email{Tommy.Nilsson@nottingham.ac.uk}

\author{Glyn Lawson}
\affiliation{%
 \institution{University of Nottingham}
  \city{Nottingham, UK}
}
\email{Glyn.Lawson@nottingham.ac.uk}

\author{Sue V.G. Cobb}
\affiliation{%
 \institution{University of Nottingham}
  \city{Nottingham, UK}
}
\email{Sue.Cobb@nottingham.ac.uk}

\author{Daniel Miller}
\affiliation{%
 \institution{University of Nottingham}
  \city{Nottingham, UK}
}
\email{Daniel.Miller@nottingham.ac.uk}

\begin{comment}
\begin{center}
  \normalsize
  Author A\textsuperscript{11}, Author B\textsuperscript{2},
  Author C\textsuperscript{1}, Author D\textsuperscript{2} and
  Author E\textsuperscript{2} \bigskip

  \textsuperscript{1}Department of Computer Science, \LaTeX\ University \par
  \textsuperscript{2}Department of Mechanical Engineering, \LaTeX\ University\par \bigskip

\end{center}
\end{comment}

\begin{abstract}
Understanding validity of user behaviour in Virtual Environments (VEs) is critical as they are increasingly being used for serious Health and Safety applications such as predicting human behaviour and training in hazardous situations. This paper presents a comparative study exploring user behaviour in VE-based fire evacuation and investigates whether this is affected by the addition of thermal and olfactory simulation. Participants (N=43) were exposed to a virtual fire in an office building. Quantitative and qualitative analyses of participant attitudes and behaviours found deviations from those we would expect in real life (e.g. pre-evacuation actions), but also valid behaviours like fire avoidance. Potentially important differences were found between multisensory and audiovisual-only conditions (e.g. perceived urgency). We conclude VEs have significant potential in safety-related applications, and that multimodality may afford additional uses in this context, but the identified limitations of behavioural validity must be carefully considered to avoid misapplication of the technology.
\end{abstract}

\copyrightyear{2019} 
\acmYear{2019} 
\setcopyright{acmcopyright}
\acmConference[CHI 2019]{CHI Conference on Human Factors in Computing Systems Proceedings}{May 4--9, 2019}{Glasgow, Scotland UK}
\acmBooktitle{CHI Conference on Human Factors in Computing Systems Proceedings (CHI 2019), May 4--9, 2019, Glasgow, Scotland UK}
\acmPrice{15.00}
\acmDOI{10.1145/3290605.3300856}
\acmISBN{978-1-4503-5970-2/19/05}

\maketitle
\section{Introduction}

Urban fire is one of the leading causes of death and property damage \cite{navitas2014improving}. During the financial year 2017/18, England alone recorded over 167,000 fire-related incidents, resulting in the death of 334 individuals, injury of over 3000, and huge costs to the economy \cite{office_2018}. In combination with high profile incidents such as the Grenfell Tower fire \cite{mckee2017grenfell}, it is no surprise that awareness and interest in fire safety is rising \cite{by_2017}. 

In order to minimise the losses resulting from any future incidents, it is vital that people are aware of the appropriate actions to take in the event of a fire, including how to escape safely. To address this, a number of studies have developed various fire safety training methods designed to increase public fire safety knowledge and improve people’s responses to fire \cite{huseyin2006fire}, while Burke et al. \cite{burke2010workplace} have argued for the need to involve a broader range of disciplines and approaches to maximise the effectiveness of such interventions. 

Advances in the versatility of virtual reality (VR), enabling high fidelity visualisation and simulation of practically any environment has proven to be very useful in training, particularly when scenarios would be too dangerous in the real world \cite{hancock2008human}. For example, Tate et al. \cite{tate1997virtual} found that immersive virtual environments (VEs) constitute an effective training tool for ship-board fire-fighting, while Mól et al.\cite{mol2009virtual} demonstrated viability of VEs in simulating emergency evacuations. Other research has demonstrated the potential of VEs as tools to predict human behaviour in emergency situations\cite{lawson2011predicting}. Behaviour patterns revealed in VEs can also inform changes to the design and layout of buildings; for example, VR has been used to compare different types of emergency exit signage on way-finding and evacuation times\cite{tang2009using}.

To better understand the efficacy of VEs in these contexts, it is vital to assess the degree to which behaviour exhibited by their users corresponds to, or deviates from, the real world\cite{lawson2011predicting}. Research has uncovered potential inconsistencies undermining the credibility of VEs. Smith \& Trenholme \cite{smith2009rapid}, for instance, found that although evacuation patterns in VEs were similar to those in real life, the duration of evacuation was generally longer. Moreover, users in VEs appeared to be prone to illogical behaviour, such as opening doors with smoke coming from underneath them. Smith \& Trenholme attribute this to the absence of sensory elements that would be present in a real fire situation, such as heat or smoke scent. Chalmers and Ferko\cite{chalmers2008levels} echo this, stressing that in order to achieve valid user behaviour, it is necessary to go beyond traditional audiovisual experiences and deliver an appropriate level of stimulation for all senses. Despite these recommendations, there remains a lack of research on the impact of multisensory simulation on user behaviour in the context of fire safety VEs. In particular, research has not yet assessed whether, or how, multisensory simulation affects validity of user behaviour. We believe exploring this topic would help developers produce more effective VE safety solutions, and in turn contribute to the mitigation of adverse consequences in future real-world incidents. 

For this purpose, we have developed a VE simulating a fire situation in an office and subsequently carried out a comparative mixed methods study with 43 participants in which we explore the influence of thermal and olfactory simulation on user behaviour. Our findings suggest that simulation of the additional senses affects users in potentially important ways, such as by increasing the perceived time pressure, and reducing gaming or rule-based attitudes. We begin to assess the impact of these psychological aspects on the observable behaviours exhibited in the VE, such as proximity to fire, pre-evacuation behaviour and route choices, and draw comparisons to behaviours reported in real fire situations. We have likewise exposed a number of instances where, in spite of increased fidelity of the VE, user behaviour and attitudes still deviated from those normally exhibited in real life. This included behaviours directly resulting from the use of a VE interface (such as overshooting movements), but also behaviours resulting from the lack of connection with the VE, such as fewer pre-evacuation delays than in the real world. We conclude that understanding and factoring in such limitations of behavioural validity is crucial if VEs are to be effectively utilized in safety-related applications.

\section{Related work}
\subsection{Multimodal VEs for fire simulation}

VEs with visual displays have frequently been used in fire simulation\cite{tate1997virtual, smith2009rapid} demonstrating the value of VE to provide safe experiences of dangerous situations. Previous research has explicitly evaluated the validity of user behaviour in virtual fire scenarios, but not with multisensory feedback \cite{lawson2011predicting}. Previously, researchers have used simulated models of flames and smoke to increase the realism within VE\cite{cha2012virtual}, but this does not provide the experiential simulation of feeling the heat or smelling the smoke.

The use of haptic technology in VEs has demonstrated the effectiveness of ‘feel’ sensation during evacuation scenarios. For example, Jiang et al.\cite{jiang2005reducing} found that, in a building evacuation scenario, participants made fewer procedural errors and completed some tasks more rapidly with the addition of haptic, vibration feedback in the VE. Additionally, empirical studies (e.g. \cite{dinh1999evaluating}) have shown that the addition of olfactory simulation increases the sense of presence within VE. Rüppel \& Shatz\cite{ruppel2011designing} proposed that the use of heating jackets, radiators, treadmills and other devices may reduce the tendency for users to treat fire evacuation simulations as a game; they argue that gamers often make unrealistic decisions and attribute this to “super-soldier syndrome”\cite{barlow2005challenging,morrison2005proficient} in which users take risks because they feel invulnerable.

Research has been conducted into multisensory feedback in fire and other safety contexts, but not with focus on valid behaviour or effectiveness in relation to applied settings. An example in a recent study\cite{garcia2018evaluation} tested the use of a haptic vest incorporating thermal actuators with 23 participants in tasks which included approaching a virtual fire. They concluded that thermal feedback improves presence and realism but identified that the heat was not sufficiently well synchronised to events in the VE and that the stimuli lacked realism. However, this work was focused on the user experience, and the authors do not comment on the resulting behaviours. Nam et al.\cite{nam2005haptic} recognised the importance of thermal simulation in VE and proposed a thermal interface for delivering multisensory feedback. However, their technology was developed for hand-held devices, rather than whole-body thermal feedback as would be experienced in a fire.

\subsection{Human Behaviour in Fires}
Extensive research into real world incidents, survivor interviews, predictive analyses and studies of fire drills have been conducted over several decades to provide a comprehensive understanding of Human Behaviour in Fires (HBiF), which is already well summarised in reviews (e.g. \cite{cheng2018human, kobes2010building}). It is not the intention of this paper to add to the understanding of HBiF as a whole, but rather to advance the effective use of VE in this field. Relevant aspects of HBiF which influenced our study design and VE development, and upon which we interpret our results, include: 

\textbf{Exit routes: }In real fire evacuation situations, people tend to evacuate via a known route, usually the one they entered through, and usually via the main entrance/exit to the building. They may not use the optimum route or evaluate all alternatives\cite{edelman1980model, shields2000study, ozel2001time, pan2006human, jeon2009characteristic, xudong2009study}. People have been reported to use lifts in evacuation\cite{proulx2006occupant}.

\textbf{Risk perception: }People with higher perceived risk respond more quickly or are more likely to leave. The level of action taken is appropriate to their perception of the environment\cite{edelman1980model, wood1980survey, galea2009uk}. People who disregard an emergency have a lower perception of risk\cite{mcconnell2009analysis}.

\textbf{Evacuation cues: }Sensory cues contribute to initiation of evacuation\cite{edelman1980model, wood1980survey, proulx1995evacuation, gershon2007factors}. People are more likely to define the situation as a fire when there are a higher number of fire cues, a consistent set of cues, and unambiguous cues\cite{kuligowski2009process}.

\textbf{Pre-evacuation delays: }People often complete non- evacuation activities before evacuating, such as: seeking  information, collecting belongings, taking emergency equipment, changing/clothes footwear, finding children, making phone calls, shutting down computers, securing items, or seeking permission to leave\cite{proulx1995evacuation, purser2001quantification, gwynne2003collection, proulx2006occupant, mcconnell2009analysis}.

\section{Aims}
This study aimed to investigate how people behave in VE in a fire evacuation scenario, and to see whether their perceptions and actions differ when thermal and olfactory feedback are added. The objective was to see whether the additional multisensory feedback makes participants respond more realistically to the hazardous situation presented in the VR, to overcome the validity issues earlier discussed. We aimed to analyse validity by comparing behaviours exhibited in the VE with real world behaviours, based on categorisations taken from expert analysis of real incidents \cite{lawson2011predicting, canter1980domestic}.
%\begin{comment}
\section{Formative studies of multisensory VEs}
The first stage of the project focused on exploring different technical solutions, informed by researching user requirements and potential use cases with industry partners. The hardware was selected with consideration to the practicalities of using the system for Health and Safety training in industry. Similar to prior research\cite{lawson2016future} industrial stakeholders expressed a desire for affordable hardware, which could be implemented ubiquitously rather than in a dedicated VR training centre. Thus, we investigated solutions with low-cost hardware and a small physical footprint. Variants of a functional multisensory VE prototype were developed and presented to stakeholders for further formative development. We used an HTC Vive\cite{vive} head-mounted display (HMD) to display Unity\cite{unity} VE scenes, with Vive controllers for interaction. An office building with multiple levels was chosen for the VE as there is literature available describing human behaviour in real world situations for this type of environment\cite{gershon2007factors,wood1980survey,aguirre1998test}. It was necessary for behavioural analysis that participants were given a choice of exit routes, and we designed the interior such that participants would be likely to encounter the fire but could subsequently choose an alternative route if they wanted to. We designed the layout to explore behaviours such as the tendency to evacuate using familiar routes\cite{kobes2010building}. We included features characteristic of office buildings, such as a reception on the ground floor, meeting and breakout rooms, a small kitchen area, and bathrooms.

\subsection{Implementation of Heat and Smell}
\begin{figure}[h]
\includegraphics[width=0.47\textwidth, left]{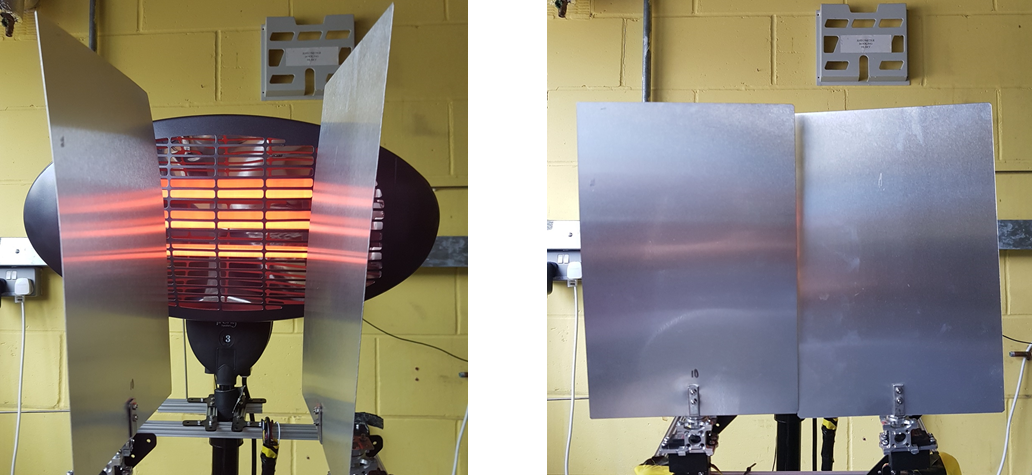}
\caption{Control of heat from IR heater: fins closed by default [right] and opening to direct heat toward user [left].}
\label{fig:heater}
\end{figure}

We conducted user studies to develop and refine the thermal and olfactory interfaces. In addition to exploring the user experience of heat and smell, this was a health and safety requirement to expose users to the minimum required level of heat and fragrance, while ensuring the sensory feedback was suitable for the purpose. Heat and smell hardware devices were controlled by an Arduino relay. For heat, IR heaters were used as the most suitable simulation of radiant heat in this application context\cite{Wareing2018user}. Initial testing with a 400-800W halogen heater showed that better control over duration and perceived direction of the heat source were required, and that heat was not strong enough. We replaced the heater with two 2KW IR heating devices in front of the user and constructed servo-controlled fins (\autoref{fig:heater}), removing synchronisation issues resulting from heaters warming up and cooling down. User studies showed that this configuration of heat sources increased perceived realism. Further, the exact location of the heaters did not impact on perceived realism: when placed symmetrically, it was not essential for the heater positions to match the relative position of the fire to the user in the VE. The timing of the heat control was found to be appropriate, and the heat was reliably perceived by all users. However, subjective feedback was that heat could be further increased to improve realism without causing discomfort, so an additional 2KW heater was subsequently added. Further details on the heat studies can be found in\cite{Wareing2018user}.

For smell, we used a Sensory Scent 200 diffuser\cite{fire} which is an electronic nebulising delivery system that can provide a concentrated, and realistic, aroma. It is compatible with an extensive selection of simulated fragrances and is a comparatively low-cost solution to scent delivery. However, these fragrances are designed for ambient scent rather than VR interfaces, and it was difficult to ascertain safe exposure levels. We consulted with health and safety and olfaction specialists, and concluded that the following best practices should be followed: (a) use only fragrances approved by the International Fragrance Association (IFRA) to ensure they meet recognised safety standards, (b) screen any users of the olfactory interface for allergies, asthma and odour intolerances, (c) use the minimum effective concentration and duration of fragrance release. 

After testing a number of smoke smells, we chose a "burning wood" fragrance oil for the scenario. We tested this with a small sample of users (N=5) to explore delivery and timeliness of scent at different concentrations (50\%, 75\%, and 100\%) and with variations on the prototype setup. Participants were asked to indicate (1) when they could detect a smell, and (2) when the smell was no longer detectable. Fragrances were released in 3 second bursts. After each trial participants were asked what they thought the smell was. After testing, we concluded that the 100\% concentration was reliably detected and associated with the smell of a fire, while lower concentrations were either not detected or not identified as fire by some participants. We found that smell was most effectively delivered with the diffuser approximately 1 metre in front of the user, a small fan behind the diffuser to direct the fragrance, and an extractor fan behind the user to control flow and remove fragrance quickly after delivery.

\subsection{Usability Studies}

A series of usability studies guided the development of the VE. These were loosely divided into three rounds of tests (five users in each) with ongoing minor iteration, but more substantial revisions in between each round. The first two rounds used think-aloud protocol\cite{ericsson1980verbal} while users navigated the prototype VE, thus obtained feedback to refine various elements of the user experience, such as optimisation of movement. An important finding was that users were perceiving the virtual building as unoccupied or abandoned, which could influence their chosen behaviours in the fire scenario. We therefore added non-player characters (NPCs) in the background through the building and ambient office sounds corresponding with the scene, and adjusted the positioning of virtual objects (chairs, papers, boxes etc.) to show evidence of occupancy.

Anecdotally, in these think-aloud sessions, we did find that participants’ comments and actions suggested that they snapped out of “game mode” when the heat and smell were activated, corresponding with \cite{ruppel2011designing}. For example, one participant started walking towards the fire and commented that he wasn’t sure he should be doing it, but continued nonetheless. As the heat fins opened, he then said \textit{“actually it’s getting warm… now I don’t think this is a good idea”}, and promptly retreated to find an alternative route. 

For the third round of tests we did not use a concurrent think-aloud protocol, as we wanted to test the entire fire evacuation scenario with a view to studying behavioural validity, and concurrent verbal protocol risks interference with task performance during verbalisation\cite{ericsson1993protocol}, which could have implications on the behaviours exhibited in this case. Instead, participants were interviewed following use of the system. In addition to further minor revisions to improve user experience, this final round of formative studies prompted us to introduce a pseudo-task (job assessment) which provided a context to improve ecological validity, and allowed us to investigate completion of pre-evacuation tasks when an incident occurs \cite{proulx1995evacuation, proulx2006occupant, purser2001quantification, mcconnell2009analysis}. 
%\end{comment}
\section{Method}

\subsection{Design}
The independent variable in our main experiment was the sensory feedback. Participants conducted a virtual fire evacuation using one of two VR configurations: one with visual and audio feedback only (AV condition), and one with visual, audio, and additional thermal and olfactory feedback (multisensory, MS condition). A between-subjects design was used because the data during a first evacuation (when the participant is unfamiliar with the experiment, task and hypotheses, and the fire is unexpected) would not be comparable with behaviour in subsequent evacuations, particularly when investigating validity. The study received ethical approval by the University of Nottingham Faculty of Engineering ethics committee.

\subsection {Participants}
Participants were recruited with advertisements posted online and emailed directly to students, staff and associates at the University of Nottingham. Participants were screened to ensure that: (a) they were not at increased risk of simulator sickness; (b) they had no previous traumatic experience in an emergency situation; and (c) they had normal or corrected-to-normal perception across all senses. Participants were assigned to the two conditions to balance as closely as possible gender, age and gaming experience (self-reported on a 5-point scale); this was intended to counter the potential confounds of individual difference factors that we had reason to expect might affect user perceptions or behaviours based on prior research\cite{kuligowski2009process, lawson2011predicting, smith2009rapid, wood1980survey}. 
52 participants were recruited. Eight experienced sickness symptoms and sessions were ended before evacuation, and one participant's data were excluded due to technology problems. Details of the 43 participants after these exclusions are in \autoref{table:demo}. \begin{table}[h]
\renewcommand{\arraystretch}{0}
\scalebox{0.9}{
\begin{tabular}{ p{5cm} p{1.5cm} p{1.5cm}}
\toprule
\textbf{Condition} & \multicolumn{1}{l}{\textbf{AV}} & \multicolumn{1}{l}{\textbf{MS}} \\ \midrule
N & 23 & 20 \\ \midrule
 Gender distribution & 10 F; 13 M & 9 F; 11 M \\ \midrule
 Mean (SD) age in years& 28.4 (5.0) & 29.3 (6.5) \\ \midrule
 Mean (SD) gaming experience & 2.91 (1.1) & 2.95 (1.3)\\ 
\midrule
\end{tabular}}
\caption{Demographic summaries of each condition.}
\label{table:demo}
\vspace{-4mm}
\end{table}
\squeezeup
\subsection{Equipment}

\begin{figure}[h]
\includegraphics[width=0.47\textwidth, left]{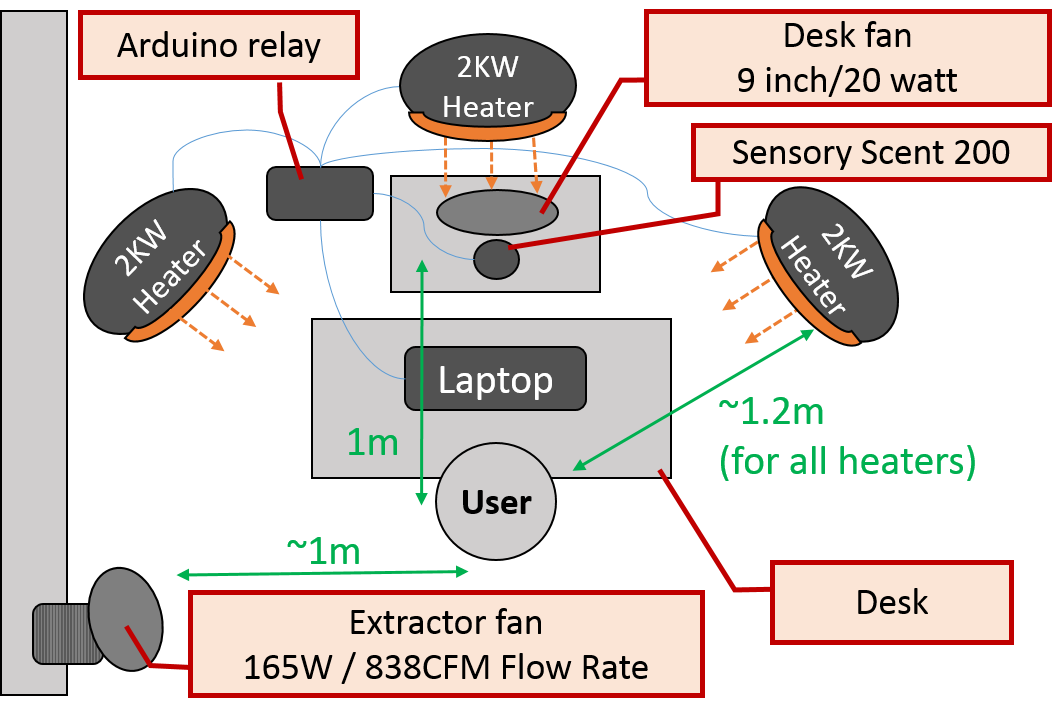}
\caption{Hardware setup after heat and smell development.}
\label{fig:setup}
\end{figure}

The VE was run on an Alienware 13 R3 laptop with external Geforce GTX 1060 graphics card. We used the Vive HMD and controllers. The right Vive control was used for turning and the left for walking movement. Movement speed could be adjusted by moving the thumb closer to or further from the centre of the control trackpad. The user could also look around the environment by moving their head. Doors opened automatically, and no interaction other than navigation was implemented. Noise cancelling headphones delivered audio simulation, which comprised background traffic (exterior) and talking (interior) sounds, as well as walking, door, and fire alarm sounds. Thermal and olfactory simulation were implemented as concluded from the studies above (\autoref{fig:setup}), controlled by Arduino, with heat delivered by three 2KW IR heaters and smell using a SensoryScent 200 fragrance diffuser with burnt wood fragrance oil. We used a desk fan and extractor to control the scent.

We used XSplitBroadcaster for screen recording and collected data logs showing positional data at each moment in the VE for analysing movement. A bespoke visualisation app was created to view the logs. A video camera was set up to capture participants’ body language and physical actions. A simulator sickness questionnaire (SSQ) \cite{kennedy1993simulator} was used to assess symptoms. A post-task questionnaire was used to assess subjective responses. 

\subsection {Procedure}

After an introduction to the experiment, participants were asked to review an information sheet and sign a consent form. They completed the SSQ before and after the study to allow monitoring of their condition and to collect information for future user studies using this system. A rapid wellbeing test\cite{graybiel1969progressive}, in which participants verbally rate on a scale from 1 (I feel fine) to 10 (I feel awful, like I will vomit), was used after the practice trial. It was also used if at any time the investigator observed indications for concern, such as a change in breathing pattern. This was a quick way to monitor as often as required without being disruptive. In cases where wellness scores worsened, sessions were terminated.
\begin{figure*}[h]
\includegraphics[width=0.92\textwidth]{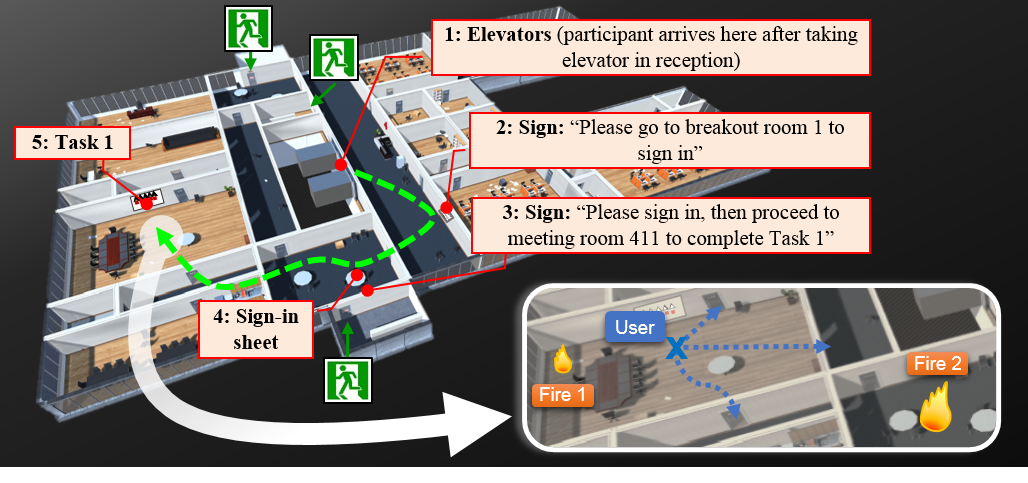}
\caption{Upper floor of the office VE showing route participants are directed pre-fire, and potential routes post-fire (inset).}
\label{fig:map}
\end{figure*} 
Participants were told that the study was investigating how people behave in VE, and were not specifically told that they would be encountering a fire scenario. They were given a practise session to become familiar with the VE and controls, in which they were instructed on how to navigate and were given a short test task in the car park outside the office building. This was typically 2-3 minutes, unless the participant required additional practice (this  was tested by the facilitator by asking them to walk from a tree to a parked car first slowly and then quickly, loop around the car, and reorientate themselves toward the building). Participants were then asked to imagine that they had been invited for an assessment for a new job that would involve some skills tests.
They were given free navigation of the VE, but were asked to follow signs within the building that would direct them to their tasks. These directed participants to an upper floor of the building via an elevator, to a breakout room to sign in, and then across a corridor to room 411, which displayed a task (an IQ test puzzle) on a large whiteboard. The route is shown in \autoref{fig:map}. This initial task sequence served the purposes of familiarising participants with the building, giving them a known entry route, and providing a distractor task.

When the participant had engaged with Task 1, the investigator manually triggered the fire scenario. This activated a small fire in room 411 (\autoref{fig:map}). In the MS condition, the fire was accompanied by release of the fragrance, and the heater fins opened to increase temperature. In both conditions, the activation also triggered the sound of a fire alarm. In addition, a second larger fire was activated in a different room in the VE, and all NPCs were teleported to an evacuation point outside the building (this was not visible to the user, as there were no NPCs in room 411).

\begin{figure}[h]
\includegraphics[width=0.46\textwidth, center]{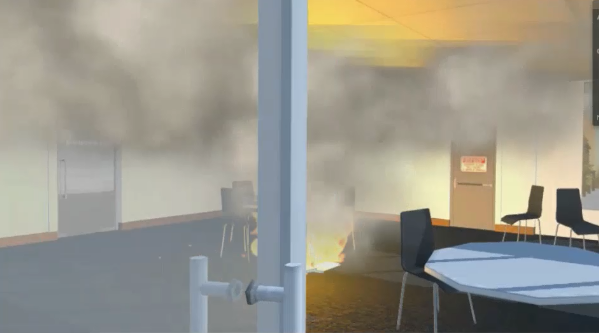}
\caption{Screenshot of scene from the office VE.}
\label{fig:smoke}
\end{figure}

The user could choose from three doors to exit room 411 (blue dotted lines show routes out of the room in \autoref{fig:map}) and there are four building exits: three marked as fire exits, and the elevators they had used earlier. Participants were not given any instruction on how to respond, but their chosen actions were recorded with screen capture and data logs. Evacuation was considered complete after participants had exited the building. The task section of the study took up to 10 minutes. \autoref{fig:smoke} shows a screenshot of the VE including the fire simulation.

After evacuation, participants were re-assessed for simulator sickness and completed the post-task subjective questionnaire. This was followed by a short unstructured interview about their thoughts and actions in the scenario, including questions about cues that made them think there was a fire, their rationale for pre-evacuation actions, evacuation route and exit choice, and whether they noticed the larger fire and what their thoughts were on encountering it. Additionally, any unusual behaviours or actions of note were explored. Participants were given the option to qualify any of their subjective ratings, or to comment on any aspect of their experience. They were then fully debriefed with the purposes of the study and asked whether they were happy to have their data included in analysis following this disclosure. In total sessions took up to 45 minutes.
\begin{table*}[t!]
\scalebox{0.9}{
\renewcommand{\arraystretch}{0}
\begin{tabular}{ p{8.5cm} p{1cm} p{1cm} p{1cm} p{1cm} p{1cm} p{1cm} p{1cm} p{1cm}}
\toprule
\textbf{Condition} & \multicolumn{2}{l}{\textbf{Median}} & \multicolumn{2}{l}{\textbf{Mean Rank}} & \multicolumn{2}{c}{\textbf{IQR}} & \multicolumn{2}{l}{\textbf{Test Statistics}} \\ \midrule
 & AV & MS & AV & MS & AV & MS & U & \textit{z}\\ \midrule
Perceived level of risk & 3 &	3 &	20.8	& 23.4 &	2 &	2 &	478 &	-0.704 \\ \midrule
Level of stress/anxiety & 3 &	3 &	20.5 &	23.7 &	2 &	2 &	472.5 &	-0.849\\  \midrule
\belowrulesepcolor{blue} \rowcolor{blue} \textbf{Level of time pressure} & 3 &	4 &	19.2 &	25.2 &	2 &	1 & \textbf{441.5*} &	\textbf{-1.646}\\ \aboverulesepcolor{blue} \midrule 
“I need to exit the building as quickly as possible” & 4 &	5 &	20.3 &	24.0 &	1 &	2 &	466 & -1.053\\  \midrule
\belowrulesepcolor{blue} \rowcolor{blue}\textbf{“The building is on fire”} & 4 &	5 &	16.7 &	28.1 &	1 &	1 &	\textbf{384*} &	\textbf{-3.112}\\ \aboverulesepcolor{blue} \midrule
\belowrulesepcolor{blue}\rowcolor{blue}\textbf{“I need to find the exit nearest to me”} & 5 &	5 &	19.4 &	25.5 &	1 &	0 &	\textbf{445*} &	\textbf{-1.837} \\ \aboverulesepcolor{blue} 
\midrule
\end{tabular}}
\caption{Results of subjective responses; highlighted rows indicate significant difference.}
\label{table:subjective}
\end{table*}
\section{Results}
\subsection{Subjective results}
Subjective results are based on participant responses to the post-task questionnaire. For the first three questions participants were asked to rate their maximum perceived level of risk, level of stress/anxiety, and level of time pressure they experienced in the scenario, where 1 = very low and 5 = very high. Participants were then given statements and were asked to indicate how true each was of their thoughts while in the VE, on a scale from 1 (not at all) to 5 (extremely).  

A Mann Whitney U-test was conducted to analyse question responses  (\autoref{table:subjective}). Participants felt a significantly higher level of time pressure in the MS (mean rank = 23.4) than AV (mean rank = 20.8) condition (one-tailed, p = .0498). Participants also had significantly higher scores for the statement “the building is on fire” in the MS (mean rank = 28.1) than AV (mean rank = 16.7) condition (one-tailed, p = .001), and significantly higher agreement scores for the statement “I need to take the nearest exit” in the MS (mean rank = 25.5) than AV (mean rank = 19.4) condition (one-tailed, p = .03).

\subsection{Evacuation times and behaviours}
Based on the HBiF literature described earlier\cite{kuligowski2009process} related to impact of cues on evacuation, we investigated the effect of multisensory feedback on evacuation times. We present this (\autoref{table:evacuation}) as follows: pre-evacuation (for which we used the time that had elapsed before participant exited the first room); evacuation time (how long it took the participant to get out of the building after initiating evacuation), and total egress time (duration from start to finish). 

\begin{table}[h]
\renewcommand{\arraystretch}{0}
\scalebox{0.9}{
\begin{tabular}{ p{1.8cm} p{0.8cm} p{0.8cm} p{0.8cm} p{0.8cm} p{0.8cm} p{0.8cm}}
\toprule
\textbf{ } & \multicolumn{2}{l}{\textbf{Pre-evacuation}} & \multicolumn{2}{l}{\textbf{Evacuation}} & \multicolumn{2}{l}{\textbf{Total egress}} \\ 
\midrule
  & MS & AV & MS & AV & MS & AV \\
 \midrule
Mean time & 23 & 23 & 44 & 40 & 67 & 60 \\
\midrule
SD & 14 & 10 & 24 & 21 & 35 & 24\\
\midrule
\end{tabular}}
\caption{Mean evacuation times (s) for each condition.}
\label{table:evacuation}
\vspace{-4mm}
\end{table}

Note that this excludes three participants (2 MS, 1 AV) who did not respond to the fire cue at all and were eventually prompted by the investigator verbally to “act as you think you would if you were really in the building”. These pre-evacuation times were 80, 86 and 69 seconds respectively. 

The results are similar between the two conditions, with slightly higher times for the MS than the AV condition. This finding contradicts the literature which suggests that additional cues would prompt faster initiation of evacuation. However, analysis of behaviour showed that times were affected by several factors that make interpretation complex. 

With the exception of the three participants who required prompts, participants spent anywhere from 10 to 57 seconds before exiting the room with the first fire. The ways in which participants responded were variable. Very few participants tried to continue with the task, and those who did stopped when they observed a sign of the fire (the sight of smoke, and/or the heat and smell). However, some participants attempted to take actions to investigate or fight the fire. For example, one participant walked up to the fire, then walked around and said aloud, “\textit{Where is there a fire extinguisher?}”. Another participant walked towards the fire and physically shook the Vive controllers at it, in an attempt to see if he could put it out. 74\% of MS and 70\% of AV participants at least investigated the situation prior to beginning evacuation.

Similar behaviours also affected evacuation time (post-fire) for some participants; for example, one participant walked through to the kitchen because they were looking for a source of water to fight the fire. Some participants also got lost while searching for an exit; this increased egress times substantially. Generally, evacuation times were most affected by the route participants chose to exit the building, and whether they changed route after encountering the second fire.

\subsection{Route choice and proximity to fire}

We deliberately gave users a known entry route by starting them outside the building and giving them a series of signs/tasks through the corridors and rooms so that we could compare this to their chosen exit route. We found that 78\% of participants attempted to retrace their entry route after the fire trigger. This shows that users demonstrate the same tendency in the VE as in the real world, and this was similar in both conditions (80\% in MS, 76\% in AV).

\begin{table}[h]
\renewcommand{\arraystretch}{0}
\scalebox{0.9}{
\begin{tabular}{  p{0.6cm} p{0.6cm} p{0.6cm} p{6.2cm}}
\toprule
\textbf{\#} & \textbf{N (MS)} & \textbf{N (AV)} & \textbf{Category Description} \\ 
\midrule
\textbf{1} & 3 & 3 & Did not encounter at all \\ 
\midrule
\textbf{2} & 7 & 5 & Witnessed fire through door and did not open door \\ 
\midrule
\textbf{3} & 3 & 2 & Moved to entrance and opened door, but subsequently did not enter room \\ 
\midrule
\textbf{4} & 1 & 3 & Entered room, but exited without attempting to cross the room \\ 
\midrule
\textbf{5} & 3 & 1 & Entered and crossed the room, with clearly visible attempt to avoid fire \\ 
\midrule
\textbf{6} & 2 & 5 & Entered and crossed the room, with no visible attempt to avoid fire \\ 
\midrule
\end{tabular}}
\caption{Categorisation of proximity to the second fire.}
\label{table:categorisation}
\vspace{-4mm}
\end{table}

\begin{table*}[h!]
\renewcommand{\arraystretch}{0}
\scalebox{0.9}{
\begin{tabular}{ p{0.8cm} p{5cm} p{9.2cm} p{1.5cm} p{1.5cm}}
\toprule
\textbf{Code} & \textbf{Behaviour category} & \textbf{Examples} & \textbf{\% MS} & \textbf{\% AV} \\ \midrule

A & Evasive &	Stepping back from fire; changing path to avoid fire & 90 & 90 \\ \midrule

B &	Wait &	Hesitation or pause &	69 &	55 \\ \midrule

C &	Seek information and investigate & Looking for extinguisher; exploring room before evacuating &	74 &	70 \\ \midrule

D &	Leave immediate area & Moving away from immediate location of fire after initiated &	68 & 70 \\ \midrule

E &	Change plan &	Changing the direction of walking after experiencing stimulus &	53 &	55 \\ \midrule

F &	Overshoot &	Walking past a door, exit or other object; colliding with a wall &	68 & 55 \\ \midrule

G &	Stay calm & Continue to act calmly as before the incident & 21 &	55 \\ \midrule

H &	Disregard/ignore stimulus &	Continue an activity unrelated to fire; walk through fire & 16 & 35 \\ \midrule

I &	Unambiguous reference to stimulus & Specific mention of a stimulus in relation to evacuation & 84 &	30 \\ \midrule

J &	Express negative feelings & Commenting on fear, anxiety, stress, panic &	63 & 5\\ \midrule

K* &	Startled reaction &	Verbal indications (e.g. “oh!”); gasps; physically jumping &	37 &	5\\ \midrule

L* &	Urgent movement &	Faster or more erratic movements &	84 &	60\\ \midrule

M* &	Attempt direct interaction &	Attempting to physically interact with the environment or objects in it, e.g. waving hands/controllers at the fire &	16 &	0\\ \midrule

N* &	Inspect escape route &	Looking through or peering around doors &	63 &	45\\ \midrule
\end{tabular}}
\caption{Categorisation of VE behaviours. Asterisk indicates those not referenced in \cite{canter1980domestic}'s taxonomy.}
\label{table:categorisation2}
\vspace{-4mm}
\end{table*}

The second fire was placed in the breakout room, partially obstructing the known route the participants had entered through, on the assumption that a majority of participants would try to evacuate this way so that we could see how they responded. We analysed how close participants got to this second fire in the VE during their evacuation, as previous research shows that in VE users do not respond realistically to risk, as earlier discussed. Behaviour around the second fire hazard was analysed from screen capture and data logs, and has been grouped into six categories (\autoref{table:categorisation}).

\begin{figure} [b]
\includegraphics[width=0.48\textwidth, left]{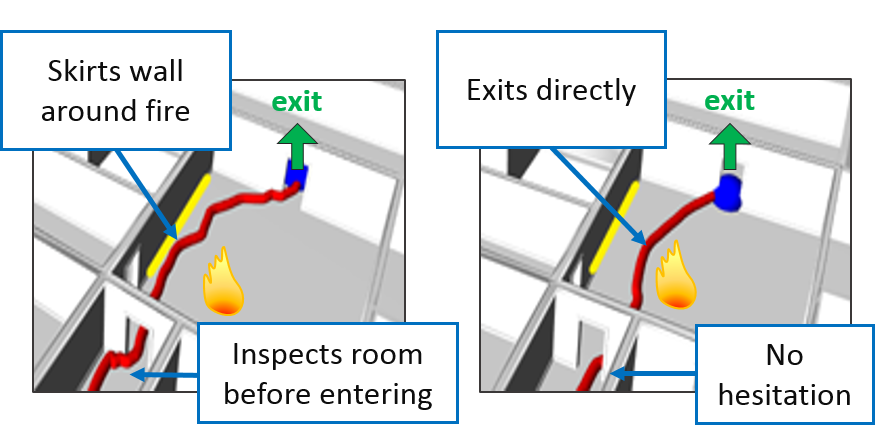}
\caption{Example paths through room (red line). Left: Category 5 (MS participant); Right: Category 6 (AV participant).}
\label{fig:category5}
\end{figure}

The majority of participants took steps to avoid the second fire. They either did not enter the room at all or entered and subsequently retreated without crossing the room. 11 participants attempted to go through the room to either the main exit (that they had come in through earlier) or the fire exit visible behind the fire. Among these participants there appears to be a distinction between the two conditions: those in MS tended to make a visible attempt to avoid the fire (category 5), such as by edging around the wall, while those in AV tended to walk very close to or through the fire (category 6) with minimal deviation or hesitation. Examples of user movement are shown in \autoref{fig:category5}; the red line indicates the user's path from positional data. In the left image the user waits outside the room peering through the door, and moves immediately to the wall before rapidly moving towards the door. In the right image, the user walks almost directly from door to door, in a smooth movement without hesitation.

\subsection {Other behavioural observations}
In addition to the measures already described, we analysed emergent behaviours to see what could be interpreted from VE behaviour. This analysis was exploratory rather than planned \textit{a priori}. The following observations were based on a combination of screen capture, data logs, video camera footage and interview responses. 
Using a similar approach to \cite{lawson2011predicting}, we reviewed VE behaviour against reference studies of behaviours from the real world. Of behaviours identified in the VE (\autoref{table:categorisation2}) the majority (codes A-J) are evident in literature on real fire situations, and are referenced in a taxonomy of acts\cite{canter1980domestic}. In other cases, the VE elicited responses that are not evident in real incidents but which are useful to understanding behaviour.

\section{Discussion}
\subsection {Comparison of MS and AV interfaces}
Results suggest that the participants in the MS condition felt a greater sense of urgency compared to those in the AV condition, as evidenced by their ratings of time pressure and their increased agreement with the statement “take the nearest exit”. This is also evidenced in their behaviour, with participants tending to exhibit behaviours reflecting urgency after encountering the fire such as erratic and quick movement, which was markedly different from the way they explored the VE before the fire was activated. This was also observed, though less, in the AV condition.

Results showed no statistical difference in participant subjective ratings of perceived risk or stress/anxiety. This is an interesting finding; while intuitively we may expect the perceived risk to be higher in MS, there are a number of factors that may influence this, some of which are evident from participant interviews after the task. Some participants qualified that they knew they were not really at risk, but felt they would be if in the real situation. Participants rating perceived risk may have been generally considering how at risk they felt, or how risky the situation theoretically was. While the difference in stress/anxiety scores was not significant, MS participants explicitly mentioned negative feelings (such as fear, stress or panic) and showed observable startled reactions more than AV participants, which is indicative of increased emotional response. This was often very visible in the video camera footage; for example, one participant almost dropped the Vive controllers, while others gasped or shouted “\textit{oh no!}”.

There was a statistical difference in agreement with the statement “The building is on fire”, with participants in MS agreeing more strongly with this statement than participants in AV. This was a particularly large effect (z=-3.112). This was also implicit in the language people used, for example “\textit{the only problem I had was [in] the room that was on fire}” (MS participant; emphasis ours). According to Kuligowski \cite{kuligowski2009process}, an increase in consistent and unambiguous cues (which is evident in our finding that people in MS made specific mention to sensory stimuli while those in AV did not) makes people more likely to interpret the situation as a fire, which logically links to this finding.

So while the reflectively-reported perceived risk did not differ, there is evidence that participant attitudes to the fire were implicitly different between the conditions. Subjective responses may have been influenced by question framing. Where participants were asked to reflect on their real psychological state (I was at risk/I was anxious) there were no differences, but for questions phrased more objectively (the building is on fire) differences were observed. 

Meanwhile, there are several examples from interviews where AV participants refer to the VE as if they are playing a game, for example: “\textit{I did not think of doing anything else because I cannot do anything else in this environment}” or, “\textit{I felt like I was in a game, so for example there were some obstacles that I should have probably gone around, but I just went through.}” Despite no differences except in sensory feedback, and therefore having the same interaction constraints and affordances that would lead to these comments, the MS condition produced far fewer examples of this. In MS participants describe a more direct and emotional interaction with the VE, for example: “\textit{I felt calmer when I got to the stairs because there was no fire there, it was quite scary}”. This may be related to our finding that some MS participants interacted directly with the VE, while AV interacted only with the prescribed interface, supporting the interpretation that MS felt more present in the environment while AV felt they were in a simulation. If participants are more detached from the experience and perceive it as a game, we suggest this increases the likelihood that in responding to the subjective questionnaire participants were considering "my avatar was at serious risk" rather than "I was at serious risk", which would account for variability in scores described earlier.  

Furthermore, differences were also observed in the way users actually behaved in the VE. Analysis of behaviour around the second fire shows differences in how people responded to the fire risk. While a similar proportion of participants in both conditions entered the room, the majority in the MS condition actively avoided the fire while the majority in the AV condition did not, in some cases even walking directly through the flames. This is clearly an unrealistic behaviour that people would be unlikely to exhibit in real life. Since not all participants went through the room, there is an insufficient sample to statistically compare this, but qualitative analyses suggest that the participants were more anxious about encountering the fire in the MS than AV condition. Future studies could force exit past a fire hazard to explore this further.

A notable difference between MS and AV was in specific unambiguous mentions of a sensory stimulus in relation to evacuation (category I). Example comments included: “\textit{it was the smell that made me think, ok […] I would run away}”; “\textit{the smell made me feel really uncomfortable and I wanted to get out fast}”; “\textit{oh god it’s smelling as well}”; and “\textit{the heat in the second fire was particularly stressful, so I looked for another way out}”. Of the participants in the MS condition, 5 explicitly commented on the heat, 9 on the smell, and 2 on both heat and smell.
Overall, in both conditions participants exhibited behaviours seen in real-world fires, for example: investigating the fire, and changing their plan after encountering a stimulus. However it is our opinion that where participants in MS seemed to show adaptive behaviour to what they were experiencing within the VE, participants in AV appeared to apply logic relating to schemas on fire evacuation, fire behaviour and building layout. The apparent distinction is in whether people were considering a projection of what they would do if this were a real situation, as opposed to responding to their direct experience. Future research could use different subjective ratings to explicitly explore this, comparing the risk to the self and risk to the projected self, with statements such as “If this were a real fire I would have been at risk” compared with “I felt I was at risk”, or “my virtual character would have been injured” compared with “I felt I would have been injured”. 

\subsection{Relation to real-world behaviours}
Our judgements of validity are based on how the observed behaviours in our study compare with reported behaviours in the real world \cite{canter1980domestic, lawson2011predicting}. In terms of pre-evacuation behaviours, participants had varying responses to the first fire. Literature on human behaviour in real fires shows that the more serious a person considers the fire to be, the more likely that they would immediately leave the building and the less likely that they would attempt to fight the fire\cite{canter1980domestic}. High seriousness rating is correlated with objective measures such as high levels of smoke spread and density. People are more likely to attempt to fight the fire if they think there is a good chance of extinguishing it, but choosing to fight the fire, i.e. judging it not to be serious, could also reflect the layperson’s misjudgement of the seriousness of fire\cite{canter1980domestic}. Based on this we would expect there to be a difference in the way users of our VE respond once encountering the second fire in comparison to the first (which was small, and contained in a waste bin), and this may explain some of the pre-evacuation behaviours seen when participants saw the initial small fire. On encountering the second, larger fire, no participants tried to fight it or to engage in any action other than evacuate. Participant comments suggest that the perception of the level of seriousness increased once they saw the second fire, and that confirmed the decision to evacuate quickly (had they not already decided to). Thus, in the VE as in the real world, participants are more likely to engage in activities other than immediate evacuation when the risk is (perceptually) lower.
Behaviours categorised as evasive were the most common across both conditions: almost all users took active steps to avoid hazards in at least one part of evacuation.

In real life we would expect more actions that delay evacuation than were evident from VE behaviour. For example, people might save their work or stop to collect their belongings\cite{kobes2009hotel}. In our study, it was hard to replicate these motivations realistically. We could give people virtual belongings, but they would be unlikely to feel particularly attached to them. We settled for a distractor task, but there was no real penalty for failing to complete it. Few participants attempted to complete the puzzle after the fire trigger, and it is unclear whether this was due to the fire cues or due to the fact they were not sufficiently engaged in the task. We suggest that the VE is not very representative of real-world behaviour in this respect. Future studies could investigate additional incentive for completing VE tasks to see if this makes a difference to behaviour.

There are also more obvious examples of deviation from real world behaviour. For example, in real life a user would not walk through flames. The very motivation for using VE is that it does not pose genuine risk to users when exploring dangerous situations, but this lowered risk evidently does lead to some deviation in behavioural response to the VE from the real world.

In both conditions there were behaviours arising specifically from VE interaction, such as difficulties with control (e.g. accidentally walking past a door). The interface layer of the VE, in the current implementation, also means we cannot compare directly some VE behaviour with real world behaviour such as physical running. However, the VE allows exploration of acts that may be present in real-world incidents that would be otherwise difficult to capture, such as skirting round a wall and gasping; these behaviours are unlikely to be reported in incident analyses. This demonstrates another potentially beneficial application of VE - to understand the details of user actions and responses in an emergency that are difficult to understand through other means, although further work will be required to understand the validity of these actions since the current approach (comparison to behavioural taxonomy) is only effective for behaviours common to both real and virtual environments.

\subsection{Implications for VEs applied in safety contexts}
Based on our findings, we are optimistic about the use of VE in safety contexts. There are many examples of user interaction in and with the VE that could be beneficial in serious applications.

The difference in psychological state with the addition of multisensory feedback may be of particular importance in training applications, as it could further afford state-dependent memory \cite{lang2001fear}; information and actions learned through the multisensory interface may be better remembered in the real situation. Therefore we recommend the use of multisensory interfaces in safety training, but suggest that research to evaluate the effectiveness of MS VE safety training would be a beneficial next step. 

The psychological distinction may be less persuasive for the use of MS in predictive applications; in these cases, the ultimate actions of the users rather than the psychological experience is likely to be most important. Predictions of human behaviour based on VE simulations should be treated with caution where behaviours are invalid, such as VE users’ likelihood of taking risks with fire. In addition to inherently risky behaviour, it is hard to represent the intrinsic motivation to save work or personal belongings in the VE, so caution should be taken interpreting behaviour around pre-evacuation delays. Any use of VE as a predictive tool should also take into account the discrepancies arising from the VE interface itself. However, pre-evacuation behaviour in terms of actions around assessment of the fire and decisions on firefighting may be valid. VE behaviour (whether multisensory or not) also appears to correspond with real-world behaviour in terms of exit choice (e.g. going back through the known route), which may be useful when considering, for example, emergency signage in a real building. VEs can therefore support accurate prediction in related areas, such as design of building layout for safe evacuation accounting for HBiF.

Our qualitative exploration of user behaviour gives an indication of the types of behaviours that can be studied within a VE, as well as an indication of where multisensory simulation may make a notable difference. The findings from this are exploratory, and can inform the design of future experiments for full inferential analysis of behaviours. As a next step towards supporting the application of VE in fire safety, we aim to formalise our framework for analysing human behaviour in virtual fires. This would build on the behaviours of interest identified here, to refine and validate a full coding scheme for future research.

\subsection{Limitations and research opportunities}
\textit{Individual and interaction effects: }This work investigated the addition of heat and smell to the fire simulation. We did not, however, compare heat with smell, nor did we separate vision and audio. Future work should investigate the comparative effects of and interactions between the senses. Chalmers et al. \cite{Chalmers2009} note that despite the important role of multiple senses on perception, achieving accurate simulation across all senses is well beyond current technological capabilities. They argue that in designing for human perceptual systems we can adapt the fidelity of various sensory feedback channels depending on the salience of those senses to the user in a given scenario. Thus, work is required to determine the appropriate level of fidelity for each sense including any cross-modal effects \cite{Chalmers2009}.  It is known that such cross-modal effects exist in similar contexts to this; for example, Weir et al.\cite{weir2013burnar} found that some users perceived a warming of their hands when visually augmented with virtual fire and smoke. Based on explicit user comments, our data suggest that smell was more salient than heat in this scenario, but different users commented on different sensory elements, so it is likely that both have perceptual importance depending on the user and their task, as proposed by \cite{Chalmers2009}.

\textit{Level of risk:}
In our study, we consciously decided to keep the risk level moderate to reduce potential distress, as have other previous researchers (e.g. removing NPCs to avoid distress of users seeing others trapped \cite{smith2009rapid}). A requirement for our analyses was that the fire was unexpected, which prevented full disclosure of what would happen in the session; we opted instead to screen participants for previous experience with emergency to avoid triggering serious traumatic response. Future studies could see if a situation with increased risk (i.e. one in which the user would definitely have been in danger if it were real) produces different results.

\textit{Qualitative methodology: }
Post-hoc questionnaires and interviews may result in over-rationalisation of behaviour. However, alternative methods such as concurrent think-aloud protocol would break immersion so are incompatible with the aims of this study. Future studies should consider the most appropriate methodology for their respective objectives.

\textit{Social interaction: }
This study is lacked a social or collaborative element. While we did have NPCs within the VE pre-activity, we did not attempt to incorporate other people into the evacuation process and users were unable to interact with others. Inclusion of social interaction was beyond the scope of the current study. However, certain real-world behaviours in response to a fire scenario are clearly aligned to the presence of other people during evacuation (e.g. information seeking and sharing with others, maintaining groups, following others’ routes, and crowding in stairwells affecting route choice and exit times\cite{kobes2009hotel, kobes2010building}). For training applications, some simple extensions of the current VE could be applied; for example NPCs with limited interaction, so that trainees can be reminded to warn others to evacuate. For predictive tools, however, the presence of other people would have far more complex influences. Potential solutions could include a combination of multi-user VEs, AI-controlled individuals, and computational modelling of crowd behaviour.

\section{conclusion}
Virtual Environments afford a high fidelity experience of dangerous situations as compared with other simulations, demonstrations and drills, while being safe and practical. As previously suggested in the literature, people do respond to fire in VE in many of the same ways as they do in real life. Our results show that we can further increase the validity of user behaviour in a VE simulation of a hazardous scenario by using multisensory interfaces. 
The key difference appears to be that users showed different attitudes toward the fire scenario, with MS participants feeling as though they were in a fire, and AV participants feeling as though they were in a fire simulation, with a more detached mentality. This was evidenced in their actions, descriptions and subjective ratings. In essence, it is the difference between “the cues tell me this is a fire scenario and I know the rules I need to apply to the game so I’ll evacuate” versus “the building is on fire and I need to evacuate”. When the application concerns life or death situations, you want people to feel like the building is on fire.

%\begin{figure}
%\includegraphics[height=1in, width=1in]{fly}
%\caption{A sample black and white graphic
%that has been resized with the \texttt{includegraphics} %command.}
%\end{figure}

\begin{acks}
This work was funded by the Institute of Occupational Safety and Health through their research fund. We gratefully acknowledge Tony Glover of DRT software for work on the VE and data visualisation app. We thank Jocelyn Spence for her helpful review, and the CHI reviewers and ACs whose suggestions have improved this paper.

\end{acks}

\bibliographystyle{ACM-Reference-Format}
\bibliography{sample-bibliography}


\begin{thebibliography}{00}

%%% the Web address.

\ifx \showCODEN    \undefined \def \showCODEN     #1{\unskip}     \fi
\ifx \showDOI      \undefined \def \showDOI       #1{#1}\fi
\ifx \showISBNx    \undefined \def \showISBNx     #1{\unskip}     \fi
\ifx \showISBNxiii \undefined \def \showISBNxiii  #1{\unskip}     \fi
\ifx \showISSN     \undefined \def \showISSN      #1{\unskip}     \fi
\ifx \showLCCN     \undefined \def \showLCCN      #1{\unskip}     \fi
\ifx \shownote     \undefined \def \shownote      #1{#1}          \fi
\ifx \showarticletitle \undefined \def \showarticletitle #1{#1}   \fi
\ifx \showURL      \undefined \def \showURL       {\relax}        \fi
% invisible to TeX
\providecommand\bibfield[2]{#2}
\providecommand\bibinfo[2]{#2}
\providecommand\natexlab[1]{#1}
\providecommand\showeprint[2][]{arXiv:#2}

\bibitem[\protect\citeauthoryear{Aguirre, Wenger, and Vigo}{Aguirre
  et~al\mbox{.}}{1998}]%
        {aguirre1998test}
\bibfield{author}{\bibinfo{person}{Benigno~E Aguirre}, \bibinfo{person}{Dennis
  Wenger}, {and} \bibinfo{person}{Gabriela Vigo}.}
  \bibinfo{year}{1998}\natexlab{}.
\newblock \showarticletitle{A test of the emergent norm theory of collective
  behavior}. In \bibinfo{booktitle}{{\em Sociological Forum}},
  Vol.~\bibinfo{volume}{13}. Springer, \bibinfo{pages}{301--320}.
\newblock


\bibitem[\protect\citeauthoryear{Barlow and Morrison}{Barlow and
  Morrison}{2005}]%
        {barlow2005challenging}
\bibfield{author}{\bibinfo{person}{Michael Barlow} {and} \bibinfo{person}{Peter
  Morrison}.} \bibinfo{year}{2005}\natexlab{}.
\newblock \showarticletitle{Challenging the super soldier syndrome in 1st
  person simulations}. In \bibinfo{booktitle}{{\em Proceedings of SimTecT 2005
  Conference, Sydney, Australia}}. Citeseer.
\newblock


\bibitem[\protect\citeauthoryear{Burke and Signal}{Burke and Signal}{2010}]%
        {burke2010workplace}
\bibfield{author}{\bibinfo{person}{Michael~J Burke} {and}
  \bibinfo{person}{Sloane~M Signal}.} \bibinfo{year}{2010}\natexlab{}.
\newblock \showarticletitle{Workplace safety: A multilevel, interdisciplinary
  perspective}.
\newblock In \bibinfo{booktitle}{{\em Research in personnel and human resources
  management}}. \bibinfo{publisher}{Emerald Group Publishing Limited},
  \bibinfo{pages}{1--47}.
\newblock


\bibitem[\protect\citeauthoryear{Canter, Breaux, and Sime}{Canter
  et~al\mbox{.}}{1980}]%
        {canter1980domestic}
\bibfield{author}{\bibinfo{person}{David Canter}, \bibinfo{person}{John
  Breaux}, {and} \bibinfo{person}{Jonathan Sime}.}
  \bibinfo{year}{1980}\natexlab{}.
\newblock \showarticletitle{Domestic, multiple occupancy, and hospital fires}.
\newblock \bibinfo{journal}{{\em Fires and human behaviour\/}}
  (\bibinfo{year}{1980}), \bibinfo{pages}{117--136}.
\newblock


\bibitem[\protect\citeauthoryear{Cha, Han, Lee, and Choi}{Cha
  et~al\mbox{.}}{2012}]%
        {cha2012virtual}
\bibfield{author}{\bibinfo{person}{Moohyun Cha}, \bibinfo{person}{Soonhung
  Han}, \bibinfo{person}{Jaikyung Lee}, {and} \bibinfo{person}{Byungil Choi}.}
  \bibinfo{year}{2012}\natexlab{}.
\newblock \showarticletitle{A virtual reality based fire training simulator
  integrated with fire dynamics data}.
\newblock \bibinfo{journal}{{\em Fire Safety Journal\/}}  \bibinfo{volume}{50}
  (\bibinfo{year}{2012}), \bibinfo{pages}{12--24}.
\newblock


\bibitem[\protect\citeauthoryear{Chalmers, Debattista, and
  Ramic-Brkic}{Chalmers et~al\mbox{.}}{2009}]%
        {Chalmers2009}
\bibfield{author}{\bibinfo{person}{Alan Chalmers}, \bibinfo{person}{Kurt
  Debattista}, {and} \bibinfo{person}{Belma Ramic-Brkic}.}
  \bibinfo{year}{2009}\natexlab{}.
\newblock \showarticletitle{Towards high-fidelity multi-sensory virtual
  environments}.
\newblock \bibinfo{journal}{{\em The Visual Computer\/}} \bibinfo{volume}{25},
  \bibinfo{number}{12} (\bibinfo{date}{20 Aug} \bibinfo{year}{2009}),
  \bibinfo{pages}{1101}.
\newblock
\showISSN{1432-2315}
\showDOI{%
\url{https://doi.org/10.1007/s00371-009-0389-2}}


\bibitem[\protect\citeauthoryear{Chalmers and Ferko}{Chalmers and
  Ferko}{2008}]%
        {chalmers2008levels}
\bibfield{author}{\bibinfo{person}{Alan Chalmers} {and} \bibinfo{person}{Andrej
  Ferko}.} \bibinfo{year}{2008}\natexlab{}.
\newblock \showarticletitle{Levels of realism: From virtual reality to real
  virtuality}. In \bibinfo{booktitle}{{\em Proceedings of the 24th Spring
  Conference on Computer Graphics}}. ACM, \bibinfo{pages}{19--25}.
\newblock


\bibitem[\protect\citeauthoryear{Cheng, Liu, Chen, Namilae, Thropp, and
  Seong}{Cheng et~al\mbox{.}}{2018}]%
        {cheng2018human}
\bibfield{author}{\bibinfo{person}{Yixuan Cheng}, \bibinfo{person}{Dahai Liu},
  \bibinfo{person}{Jie Chen}, \bibinfo{person}{Sirish Namilae},
  \bibinfo{person}{Jennifer Thropp}, {and} \bibinfo{person}{Younho Seong}.}
  \bibinfo{year}{2018}\natexlab{}.
\newblock \showarticletitle{Human Behavior Under Emergency and Its Simulation
  Modeling: A Review}. In \bibinfo{booktitle}{{\em International Conference on
  Applied Human Factors and Ergonomics}}. Springer, \bibinfo{pages}{313--325}.
\newblock


\bibitem[\protect\citeauthoryear{Dinh, Walker, Hodges, Song, and
  Kobayashi}{Dinh et~al\mbox{.}}{1999}]%
        {dinh1999evaluating}
\bibfield{author}{\bibinfo{person}{Huong~Q Dinh}, \bibinfo{person}{Neff
  Walker}, \bibinfo{person}{Larry~F Hodges}, \bibinfo{person}{Chang Song},
  {and} \bibinfo{person}{Akira Kobayashi}.} \bibinfo{year}{1999}\natexlab{}.
\newblock \showarticletitle{Evaluating the importance of multi-sensory input on
  memory and the sense of presence in virtual environments}. In
  \bibinfo{booktitle}{{\em Proceedings IEEE Virtual Reality (Cat. No.
  99CB36316)}}. IEEE, \bibinfo{pages}{222--228}.
\newblock


\bibitem[\protect\citeauthoryear{Edelman, Herz, and Bickman}{Edelman
  et~al\mbox{.}}{1980}]%
        {edelman1980model}
\bibfield{author}{\bibinfo{person}{P Edelman}, \bibinfo{person}{E Herz}, {and}
  \bibinfo{person}{L Bickman}.} \bibinfo{year}{1980}\natexlab{}.
\newblock \showarticletitle{A model of behaviour in fires applied to a nursing
  home fire}.
\newblock \bibinfo{journal}{{\em Fires and human behaviour\/}}
  (\bibinfo{year}{1980}), \bibinfo{pages}{181--203}.
\newblock


\bibitem[\protect\citeauthoryear{Ericsson and Simon}{Ericsson and
  Simon}{1980}]%
        {ericsson1980verbal}
\bibfield{author}{\bibinfo{person}{K~Anders Ericsson} {and}
  \bibinfo{person}{Herbert~A Simon}.} \bibinfo{year}{1980}\natexlab{}.
\newblock \showarticletitle{Verbal reports as data.}
\newblock \bibinfo{journal}{{\em Psychological review\/}} \bibinfo{volume}{87},
  \bibinfo{number}{3} (\bibinfo{year}{1980}), \bibinfo{pages}{215}.
\newblock


\bibitem[\protect\citeauthoryear{Ericsson and Simon}{Ericsson and
  Simon}{1993}]%
        {ericsson1993protocol}
\bibfield{author}{\bibinfo{person}{Karl~Anders Ericsson} {and}
  \bibinfo{person}{Herbert~Alexander Simon}.} \bibinfo{year}{1993}\natexlab{}.
\newblock \bibinfo{booktitle}{{\em Protocol analysis}}.
\newblock \bibinfo{publisher}{MIT press Cambridge, MA}.
\newblock


\bibitem[\protect\citeauthoryear{Europe}{Europe}{2018}]%
        {fire}
\bibfield{author}{\bibinfo{person}{Back Stage~Technologies Europe}.}
  \bibinfo{year}{2018}\natexlab{}.
\newblock \bibinfo{title}{Back Stage Technologies Europe:THEMEPARK,
  ENTERTAINMENT AND MOVIE SPECIAL EFFECTS}.
\newblock   (\bibinfo{year}{2018}).
\newblock
\showURL{%
\url{http://www.back-stage-technologies.co.uk}}


\bibitem[\protect\citeauthoryear{Galea, Hulse, Day, Siddiqui, and Sharp}{Galea
  et~al\mbox{.}}{2009}]%
        {galea2009uk}
\bibfield{author}{\bibinfo{person}{Edwin Galea}, \bibinfo{person}{Lynn Hulse},
  \bibinfo{person}{Rachel Day}, \bibinfo{person}{Asim Siddiqui}, {and}
  \bibinfo{person}{Gary Sharp}.} \bibinfo{year}{2009}\natexlab{}.
\newblock \showarticletitle{The UK WTC 9/11 evacuation study: an overview of
  the methodologies employed and some analysis relating to fatigue, stair
  travel speeds and occupant response times}.
\newblock  (\bibinfo{year}{2009}).
\newblock


\bibitem[\protect\citeauthoryear{Garc{\'\i}a-Valle, Ferre, Bre{\~n}osa, and
  Vargas}{Garc{\'\i}a-Valle et~al\mbox{.}}{2018}]%
        {garcia2018evaluation}
\bibfield{author}{\bibinfo{person}{Gonzalo Garc{\'\i}a-Valle},
  \bibinfo{person}{Manuel Ferre}, \bibinfo{person}{Jose Bre{\~n}osa}, {and}
  \bibinfo{person}{David Vargas}.} \bibinfo{year}{2018}\natexlab{}.
\newblock \showarticletitle{Evaluation of Presence in Virtual Environments:
  Haptic Vest and User’s Haptic Skills}.
\newblock \bibinfo{journal}{{\em IEEE Access\/}}  \bibinfo{volume}{6}
  (\bibinfo{year}{2018}), \bibinfo{pages}{7224--7233}.
\newblock


\bibitem[\protect\citeauthoryear{Gershon, Qureshi, Rubin, and Raveis}{Gershon
  et~al\mbox{.}}{2007}]%
        {gershon2007factors}
\bibfield{author}{\bibinfo{person}{Robyn~RM Gershon},
  \bibinfo{person}{Kristine~A Qureshi}, \bibinfo{person}{Marcie~S Rubin}, {and}
  \bibinfo{person}{Victoria~H Raveis}.} \bibinfo{year}{2007}\natexlab{}.
\newblock \showarticletitle{Factors associated with high-rise evacuation:
  qualitative results from the World Trade Center Evacuation Study}.
\newblock \bibinfo{journal}{{\em Prehospital and disaster medicine\/}}
  \bibinfo{volume}{22}, \bibinfo{number}{3} (\bibinfo{year}{2007}),
  \bibinfo{pages}{165--173}.
\newblock


\bibitem[\protect\citeauthoryear{Graybiel and Reason}{Graybiel and
  Reason}{1969}]%
        {graybiel1969progressive}
\bibfield{author}{\bibinfo{person}{A Graybiel} {and} \bibinfo{person}{JT
  Reason}.} \bibinfo{year}{1969}\natexlab{}.
\newblock \showarticletitle{Progressive adaptation to Coriolis accelerations
  associated with 1-rpm increments in the velocity of the slow rotation room}.
\newblock  (\bibinfo{year}{1969}).
\newblock


\bibitem[\protect\citeauthoryear{Gwynne, Galea, Parke, and Hickson}{Gwynne
  et~al\mbox{.}}{2003}]%
        {gwynne2003collection}
\bibfield{author}{\bibinfo{person}{S Gwynne}, \bibinfo{person}{ER Galea},
  \bibinfo{person}{J Parke}, {and} \bibinfo{person}{J Hickson}.}
  \bibinfo{year}{2003}\natexlab{}.
\newblock \showarticletitle{The collection and analysis of pre-evacuation times
  derived from evacuation trials and their application to evacuation
  modelling}.
\newblock \bibinfo{journal}{{\em Fire Technology\/}} \bibinfo{volume}{39},
  \bibinfo{number}{2} (\bibinfo{year}{2003}), \bibinfo{pages}{173--195}.
\newblock


\bibitem[\protect\citeauthoryear{Hancock, Vincenzi, Wise, and Mouloua}{Hancock
  et~al\mbox{.}}{2008}]%
        {hancock2008human}
\bibfield{author}{\bibinfo{person}{Peter~A Hancock}, \bibinfo{person}{Dennis~A
  Vincenzi}, \bibinfo{person}{John~A Wise}, {and} \bibinfo{person}{Mustapha
  Mouloua}.} \bibinfo{year}{2008}\natexlab{}.
\newblock \bibinfo{booktitle}{{\em Human factors in simulation and training}}.
\newblock \bibinfo{publisher}{CRC Press}.
\newblock


\bibitem[\protect\citeauthoryear{Huseyin, Satyen, et~al\mbox{.}}{Huseyin
  et~al\mbox{.}}{2006}]%
        {huseyin2006fire}
\bibfield{author}{\bibinfo{person}{Ilmiye Huseyin}, \bibinfo{person}{Lata
  Satyen}, {et~al\mbox{.}}} \bibinfo{year}{2006}\natexlab{}.
\newblock \showarticletitle{Fire safety training: Its importance in enhancing
  fire safety knowledge and response to fire}.
\newblock \bibinfo{journal}{{\em Australian Journal of Emergency Management,
  The\/}} \bibinfo{volume}{21}, \bibinfo{number}{4} (\bibinfo{year}{2006}),
  \bibinfo{pages}{48}.
\newblock


\bibitem[\protect\citeauthoryear{Jeon and Hong}{Jeon and Hong}{2009}]%
        {jeon2009characteristic}
\bibfield{author}{\bibinfo{person}{Gyuyeob Jeon} {and} \bibinfo{person}{Wonhwa
  Hong}.} \bibinfo{year}{2009}\natexlab{}.
\newblock \showarticletitle{Characteristic features of the behavior and
  perception of evacuees from the Daegu subway fire and safety measures in an
  underground fire}.
\newblock \bibinfo{journal}{{\em Journal of Asian Architecture and Building
  Engineering\/}} \bibinfo{volume}{8}, \bibinfo{number}{2}
  (\bibinfo{year}{2009}), \bibinfo{pages}{415--422}.
\newblock


\bibitem[\protect\citeauthoryear{Jiang, Girotra, Cutkosky, and Ullrich}{Jiang
  et~al\mbox{.}}{2005}]%
        {jiang2005reducing}
\bibfield{author}{\bibinfo{person}{Li Jiang}, \bibinfo{person}{Rohit Girotra},
  \bibinfo{person}{Mark~R Cutkosky}, {and} \bibinfo{person}{Chris Ullrich}.}
  \bibinfo{year}{2005}\natexlab{}.
\newblock \showarticletitle{Reducing error rates with low-cost haptic feedback
  in virtual reality-based training applications}. In \bibinfo{booktitle}{{\em
  Eurohaptics Conference, 2005 and Symposium on Haptic Interfaces for Virtual
  Environment and Teleoperator Systems, 2005. World Haptics 2005. First
  Joint}}. IEEE, \bibinfo{pages}{420--425}.
\newblock


\bibitem[\protect\citeauthoryear{Kennedy, Lane, Berbaum, and
  Lilienthal}{Kennedy et~al\mbox{.}}{1993}]%
        {kennedy1993simulator}
\bibfield{author}{\bibinfo{person}{Robert~S Kennedy}, \bibinfo{person}{Norman~E
  Lane}, \bibinfo{person}{Kevin~S Berbaum}, {and} \bibinfo{person}{Michael~G
  Lilienthal}.} \bibinfo{year}{1993}\natexlab{}.
\newblock \showarticletitle{Simulator sickness questionnaire: An enhanced
  method for quantifying simulator sickness}.
\newblock \bibinfo{journal}{{\em The international journal of aviation
  psychology\/}} \bibinfo{volume}{3}, \bibinfo{number}{3}
  (\bibinfo{year}{1993}), \bibinfo{pages}{203--220}.
\newblock


\bibitem[\protect\citeauthoryear{Kobes, Helsloot, De~Vries, and Post}{Kobes
  et~al\mbox{.}}{2010}]%
        {kobes2010building}
\bibfield{author}{\bibinfo{person}{Margrethe Kobes}, \bibinfo{person}{Ira
  Helsloot}, \bibinfo{person}{Bauke De~Vries}, {and} \bibinfo{person}{Jos~G
  Post}.} \bibinfo{year}{2010}\natexlab{}.
\newblock \showarticletitle{Building safety and human behaviour in fire: A
  literature review}.
\newblock \bibinfo{journal}{{\em Fire Safety Journal\/}} \bibinfo{volume}{45},
  \bibinfo{number}{1} (\bibinfo{year}{2010}), \bibinfo{pages}{1--11}.
\newblock


\bibitem[\protect\citeauthoryear{Kobes, Oberij{\'e}, Groenewegen, Helsloot, and
  De~Vries}{Kobes et~al\mbox{.}}{2009}]%
        {kobes2009hotel}
\bibfield{author}{\bibinfo{person}{M Kobes}, \bibinfo{person}{Nancy
  Oberij{\'e}}, \bibinfo{person}{Karin Groenewegen}, \bibinfo{person}{I
  Helsloot}, {and} \bibinfo{person}{B De~Vries}.}
  \bibinfo{year}{2009}\natexlab{}.
\newblock \showarticletitle{Hotel evacuation at night: an analysis of
  unannounced fire drills under various conditions}. In
  \bibinfo{booktitle}{{\em Proc. of Human Behavior in Fire Symp}},
  Vol.~\bibinfo{volume}{13}. \bibinfo{pages}{15}.
\newblock


\bibitem[\protect\citeauthoryear{Kuligowski}{Kuligowski}{2009}]%
        {kuligowski2009process}
\bibfield{author}{\bibinfo{person}{Erica~D Kuligowski}.}
  \bibinfo{year}{2009}\natexlab{}.
\newblock \bibinfo{booktitle}{{\em The process of human behavior in fires}}.
\newblock \bibinfo{publisher}{US Department of Commerce, National Institute of
  Standards and Technology~…}.
\newblock


\bibitem[\protect\citeauthoryear{Lang, Craske, Brown, and Ghaneian}{Lang
  et~al\mbox{.}}{2001}]%
        {lang2001fear}
\bibfield{author}{\bibinfo{person}{Ariel~J Lang}, \bibinfo{person}{Michelle~G
  Craske}, \bibinfo{person}{Matt Brown}, {and} \bibinfo{person}{Atousa
  Ghaneian}.} \bibinfo{year}{2001}\natexlab{}.
\newblock \showarticletitle{Fear-related state dependent memory}.
\newblock \bibinfo{journal}{{\em Cognition \& Emotion\/}} \bibinfo{volume}{15},
  \bibinfo{number}{5} (\bibinfo{year}{2001}), \bibinfo{pages}{695--703}.
\newblock


\bibitem[\protect\citeauthoryear{Lawson}{Lawson}{2011}]%
        {lawson2011predicting}
\bibfield{author}{\bibinfo{person}{Glyn Lawson}.}
  \bibinfo{year}{2011}\natexlab{}.
\newblock {\em \bibinfo{title}{Predicting human behaviour in emergencies}}.
\newblock \bibinfo{thesistype}{Ph.D. Dissertation}. \bibinfo{school}{University
  of Nottingham}.
\newblock


\bibitem[\protect\citeauthoryear{Lawson, Salanitri, and Waterfield}{Lawson
  et~al\mbox{.}}{2016}]%
        {lawson2016future}
\bibfield{author}{\bibinfo{person}{Glyn Lawson}, \bibinfo{person}{Davide
  Salanitri}, {and} \bibinfo{person}{Brian Waterfield}.}
  \bibinfo{year}{2016}\natexlab{}.
\newblock \showarticletitle{Future directions for the development of virtual
  reality within an automotive manufacturer}.
\newblock \bibinfo{journal}{{\em Applied Ergonomics\/}}  \bibinfo{volume}{53}
  (\bibinfo{year}{2016}), \bibinfo{pages}{323--330}.
\newblock


\bibitem[\protect\citeauthoryear{Marketwatch}{Marketwatch}{2017}]%
        {by_2017}
\bibfield{author}{\bibinfo{person}{Marketwatch}.}
  \bibinfo{year}{2017}\natexlab{}.
\newblock \bibinfo{title}{Global Fire Safety Systems and Equipment Market
  (2017-2023): Forecast By Types, Verticals, Regions and Competitive Landscape
  - Research and Markets}.
\newblock   (\bibinfo{date}{Dec} \bibinfo{year}{2017}).
\newblock
\showURL{%
\url{http://www.marketwatch.com}}


\bibitem[\protect\citeauthoryear{McConnell, Boyce, and Shields}{McConnell
  et~al\mbox{.}}{2009}]%
        {mcconnell2009analysis}
\bibfield{author}{\bibinfo{person}{NC McConnell}, \bibinfo{person}{KE Boyce},
  {and} \bibinfo{person}{TJ Shields}.} \bibinfo{year}{2009}\natexlab{}.
\newblock \showarticletitle{An Analysis of the Recognition and Response
  Behaviours of Evacuees of WTC 1 on 9/11}. In \bibinfo{booktitle}{{\em
  Proceedings 4th International Human Behaviour in Fire Symposium}}. Citeseer,
  \bibinfo{pages}{659--670}.
\newblock


\bibitem[\protect\citeauthoryear{McKee}{McKee}{2017}]%
        {mckee2017grenfell}
\bibfield{author}{\bibinfo{person}{Martin McKee}.}
  \bibinfo{year}{2017}\natexlab{}.
\newblock \showarticletitle{Grenfell Tower fire: why we cannot ignore the
  political determinants of health}.
\newblock \bibinfo{journal}{{\em BMJ: British Medical Journal (Online)\/}}
  \bibinfo{volume}{357} (\bibinfo{year}{2017}).
\newblock


\bibitem[\protect\citeauthoryear{M{\'o}l, Jorge, Couto, Augusto, Cunha, and
  Landau}{M{\'o}l et~al\mbox{.}}{2009}]%
        {mol2009virtual}
\bibfield{author}{\bibinfo{person}{Ant{\^o}nio Carlos~A M{\'o}l},
  \bibinfo{person}{Carlos Alexandre~F Jorge}, \bibinfo{person}{Pedro~M Couto},
  \bibinfo{person}{Silas~C Augusto}, \bibinfo{person}{Gerson~G Cunha}, {and}
  \bibinfo{person}{Luiz Landau}.} \bibinfo{year}{2009}\natexlab{}.
\newblock \showarticletitle{Virtual environments simulation for dose assessment
  in nuclear plants}.
\newblock \bibinfo{journal}{{\em Progress in Nuclear Energy\/}}
  \bibinfo{volume}{51}, \bibinfo{number}{2} (\bibinfo{year}{2009}),
  \bibinfo{pages}{382--387}.
\newblock


\bibitem[\protect\citeauthoryear{Morrison, Barlow, Bethel, and
  Clothier}{Morrison et~al\mbox{.}}{2005}]%
        {morrison2005proficient}
\bibfield{author}{\bibinfo{person}{Peter Morrison}, \bibinfo{person}{Michael
  Barlow}, \bibinfo{person}{Glen Bethel}, {and} \bibinfo{person}{Scott
  Clothier}.} \bibinfo{year}{2005}\natexlab{}.
\newblock \showarticletitle{Proficient soldier to skilled gamer: Training for
  COTS success}. In \bibinfo{booktitle}{{\em SimTect 2005 Conference
  Proceedings}}. \bibinfo{pages}{91--96}.
\newblock


\bibitem[\protect\citeauthoryear{Nam, Di, Borsodi, and Mackay}{Nam
  et~al\mbox{.}}{2005}]%
        {nam2005haptic}
\bibfield{author}{\bibinfo{person}{Chang~S Nam}, \bibinfo{person}{Jia Di},
  \bibinfo{person}{Liam~W Borsodi}, {and} \bibinfo{person}{William Mackay}.}
  \bibinfo{year}{2005}\natexlab{}.
\newblock \showarticletitle{A haptic thermal interface: Towards effective
  multimodal user interface systems}.
\newblock \bibinfo{journal}{{\em Proceedings of IASTED-HCI\/}}
  (\bibinfo{year}{2005}), \bibinfo{pages}{13--18}.
\newblock


\bibitem[\protect\citeauthoryear{Navitas}{Navitas}{2014}]%
        {navitas2014improving}
\bibfield{author}{\bibinfo{person}{Prananda Navitas}.}
  \bibinfo{year}{2014}\natexlab{}.
\newblock \showarticletitle{Improving resilience against urban fire hazards
  through environmental design in dense urban areas in Surabaya, Indonesia}.
\newblock \bibinfo{journal}{{\em Procedia-Social and Behavioral Sciences\/}}
  \bibinfo{volume}{135} (\bibinfo{year}{2014}), \bibinfo{pages}{178--183}.
\newblock


\bibitem[\protect\citeauthoryear{Office}{Office}{2018}]%
        {office_2018}
\bibfield{author}{\bibinfo{person}{Home Office}.}
  \bibinfo{year}{2018}\natexlab{}.
\newblock \bibinfo{title}{Fire and rescue incident statistics: England, year
  ending March 2018}.
\newblock   (\bibinfo{date}{Aug} \bibinfo{year}{2018}).
\newblock
\showURL{%
\url{https://www.gov.uk/government/statistics/fire-and-rescue-incident-statistics-england-year-ending-march-2018}}


\bibitem[\protect\citeauthoryear{Ozel}{Ozel}{2001}]%
        {ozel2001time}
\bibfield{author}{\bibinfo{person}{F Ozel}.} \bibinfo{year}{2001}\natexlab{}.
\newblock \showarticletitle{Time pressure and stress as a factor during
  emergency egress}.
\newblock \bibinfo{journal}{{\em Safety Science\/}} \bibinfo{volume}{38},
  \bibinfo{number}{2} (\bibinfo{year}{2001}), \bibinfo{pages}{95--107}.
\newblock


\bibitem[\protect\citeauthoryear{Pan, Han, Dauber, and Law}{Pan
  et~al\mbox{.}}{2006}]%
        {pan2006human}
\bibfield{author}{\bibinfo{person}{Xiaoshan Pan}, \bibinfo{person}{Charles~S
  Han}, \bibinfo{person}{Ken Dauber}, {and} \bibinfo{person}{Kincho~H Law}.}
  \bibinfo{year}{2006}\natexlab{}.
\newblock \showarticletitle{Human and social behavior in computational modeling
  and analysis of egress}.
\newblock \bibinfo{journal}{{\em Automation in construction\/}}
  \bibinfo{volume}{15}, \bibinfo{number}{4} (\bibinfo{year}{2006}),
  \bibinfo{pages}{448--461}.
\newblock


\bibitem[\protect\citeauthoryear{Proulx}{Proulx}{1995}]%
        {proulx1995evacuation}
\bibfield{author}{\bibinfo{person}{Guylene Proulx}.}
  \bibinfo{year}{1995}\natexlab{}.
\newblock \showarticletitle{Evacuation time and movement in apartment
  buildings}.
\newblock \bibinfo{journal}{{\em Fire safety journal\/}} \bibinfo{volume}{24},
  \bibinfo{number}{3} (\bibinfo{year}{1995}), \bibinfo{pages}{229--246}.
\newblock


\bibitem[\protect\citeauthoryear{Proulx and Reid}{Proulx and Reid}{2006}]%
        {proulx2006occupant}
\bibfield{author}{\bibinfo{person}{Guylene Proulx} {and}
  \bibinfo{person}{Irene~MA Reid}.} \bibinfo{year}{2006}\natexlab{}.
\newblock \showarticletitle{Occupant behavior and evacuation during the Chicago
  Cook County Administration Building fire}.
\newblock \bibinfo{journal}{{\em Journal of fire protection engineering\/}}
  \bibinfo{volume}{16}, \bibinfo{number}{4} (\bibinfo{year}{2006}),
  \bibinfo{pages}{283--309}.
\newblock


\bibitem[\protect\citeauthoryear{Purser and Bensilum}{Purser and
  Bensilum}{2001}]%
        {purser2001quantification}
\bibfield{author}{\bibinfo{person}{Dave~A Purser} {and} \bibinfo{person}{M
  Bensilum}.} \bibinfo{year}{2001}\natexlab{}.
\newblock \showarticletitle{Quantification of behaviour for engineering design
  standards and escape time calculations}.
\newblock \bibinfo{journal}{{\em Safety science\/}} \bibinfo{volume}{38},
  \bibinfo{number}{2} (\bibinfo{year}{2001}), \bibinfo{pages}{157--182}.
\newblock


\bibitem[\protect\citeauthoryear{R{\"u}ppel and Schatz}{R{\"u}ppel and
  Schatz}{2011}]%
        {ruppel2011designing}
\bibfield{author}{\bibinfo{person}{Uwe R{\"u}ppel} {and}
  \bibinfo{person}{Kristian Schatz}.} \bibinfo{year}{2011}\natexlab{}.
\newblock \showarticletitle{Designing a BIM-based serious game for fire safety
  evacuation simulations}.
\newblock \bibinfo{journal}{{\em Advanced Engineering Informatics\/}}
  \bibinfo{volume}{25}, \bibinfo{number}{4} (\bibinfo{year}{2011}),
  \bibinfo{pages}{600--611}.
\newblock


\bibitem[\protect\citeauthoryear{Shields and Boyce}{Shields and Boyce}{2000}]%
        {shields2000study}
\bibfield{author}{\bibinfo{person}{TJ Shields} {and} \bibinfo{person}{KE
  Boyce}.} \bibinfo{year}{2000}\natexlab{}.
\newblock \showarticletitle{A study of evacuation from large retail stores}.
\newblock \bibinfo{journal}{{\em Fire Safety Journal\/}} \bibinfo{volume}{35},
  \bibinfo{number}{1} (\bibinfo{year}{2000}), \bibinfo{pages}{25--49}.
\newblock


\bibitem[\protect\citeauthoryear{Smith and Trenholme}{Smith and
  Trenholme}{2009}]%
        {smith2009rapid}
\bibfield{author}{\bibinfo{person}{Shamus~P Smith} {and} \bibinfo{person}{David
  Trenholme}.} \bibinfo{year}{2009}\natexlab{}.
\newblock \showarticletitle{Rapid prototyping a virtual fire drill environment
  using computer game technology}.
\newblock \bibinfo{journal}{{\em Fire safety journal\/}} \bibinfo{volume}{44},
  \bibinfo{number}{4} (\bibinfo{year}{2009}), \bibinfo{pages}{559--569}.
\newblock


\bibitem[\protect\citeauthoryear{Tang, Wu, and Lin}{Tang et~al\mbox{.}}{2009}]%
        {tang2009using}
\bibfield{author}{\bibinfo{person}{Chieh-Hsin Tang}, \bibinfo{person}{Wu-Tai
  Wu}, {and} \bibinfo{person}{Ching-Yuan Lin}.}
  \bibinfo{year}{2009}\natexlab{}.
\newblock \showarticletitle{Using virtual reality to determine how emergency
  signs facilitate way-finding}.
\newblock \bibinfo{journal}{{\em Applied ergonomics\/}} \bibinfo{volume}{40},
  \bibinfo{number}{4} (\bibinfo{year}{2009}), \bibinfo{pages}{722--730}.
\newblock


\bibitem[\protect\citeauthoryear{Tate, Sibert, and King}{Tate
  et~al\mbox{.}}{1997}]%
        {tate1997virtual}
\bibfield{author}{\bibinfo{person}{David~L Tate}, \bibinfo{person}{Linda
  Sibert}, {and} \bibinfo{person}{Tony King}.} \bibinfo{year}{1997}\natexlab{}.
\newblock \showarticletitle{Virtual environments for shipboard firefighting
  training}. In \bibinfo{booktitle}{{\em Virtual reality annual international
  symposium, 1997., IEEE 1997}}. IEEE, \bibinfo{pages}{61--68}.
\newblock


\bibitem[\protect\citeauthoryear{Technologies}{Technologies}{2018}]%
        {unity}
\bibfield{author}{\bibinfo{person}{Unity Technologies}.}
  \bibinfo{year}{2018}\natexlab{}.
\newblock \bibinfo{title}{Unity}.
\newblock   (\bibinfo{year}{2018}).
\newblock
\showURL{%
\url{https://unity3d.com/}}


\bibitem[\protect\citeauthoryear{Vive}{Vive}{2018}]%
        {vive}
\bibfield{author}{\bibinfo{person}{HTC Vive}.} \bibinfo{year}{2018}\natexlab{}.
\newblock \bibinfo{title}{Discover Virtual Reality Beyond Imagination}.
\newblock   (\bibinfo{year}{2018}).
\newblock
\showURL{%
\url{https://www.vive.com/}}


\bibitem[\protect\citeauthoryear{Wareing, Lawson, Abdullah, and Roper}{Wareing
  et~al\mbox{.}}{2018}]%
        {Wareing2018user}
\bibfield{author}{\bibinfo{person}{Joseph Wareing}, \bibinfo{person}{Glyn
  Lawson}, \bibinfo{person}{Che Abdullah}, {and} \bibinfo{person}{Tessa
  Roper}.} \bibinfo{year}{2018}\natexlab{}.
\newblock \showarticletitle{User Perception of Heat Source Location for a
  Multisensory Fire Training Simulation}. In \bibinfo{booktitle}{{\em Computer
  Science and Electronic Engineering (CEEC), 2018}}. IEEE,
  \bibinfo{pages}{0--0}.
\newblock


\bibitem[\protect\citeauthoryear{Weir, Sandor, Swoboda, Nguyen, Eck, Reitmayr,
  and Day}{Weir et~al\mbox{.}}{2013}]%
        {weir2013burnar}
\bibfield{author}{\bibinfo{person}{Peter Weir}, \bibinfo{person}{Christian
  Sandor}, \bibinfo{person}{Matt Swoboda}, \bibinfo{person}{Thanh Nguyen},
  \bibinfo{person}{Ulrich Eck}, \bibinfo{person}{Gerhard Reitmayr}, {and}
  \bibinfo{person}{Arindam Day}.} \bibinfo{year}{2013}\natexlab{}.
\newblock \bibinfo{booktitle}{{\em Burnar: Involuntary heat sensations in
  augmented reality}}.
\newblock \bibinfo{publisher}{IEEE}.
\newblock


\bibitem[\protect\citeauthoryear{Wood}{Wood}{1980}]%
        {wood1980survey}
\bibfield{author}{\bibinfo{person}{Peter~G Wood}.}
  \bibinfo{year}{1980}\natexlab{}.
\newblock \bibinfo{title}{A survey of behaviour in fires}.
\newblock   (\bibinfo{year}{1980}), \bibinfo{numpages}{83--95}~pages.
\newblock


\bibitem[\protect\citeauthoryear{Xudong, Heping, Qiyuan, Yong, Hongjiang, and
  Chenjie}{Xudong et~al\mbox{.}}{2009}]%
        {xudong2009study}
\bibfield{author}{\bibinfo{person}{Cheng Xudong}, \bibinfo{person}{Zhang
  Heping}, \bibinfo{person}{Xie Qiyuan}, \bibinfo{person}{Zhou Yong},
  \bibinfo{person}{Zhang Hongjiang}, {and} \bibinfo{person}{Zhang Chenjie}.}
  \bibinfo{year}{2009}\natexlab{}.
\newblock \showarticletitle{Study of announced evacuation drill from a retail
  store}.
\newblock \bibinfo{journal}{{\em Building and Environment\/}}
  \bibinfo{volume}{44}, \bibinfo{number}{5} (\bibinfo{year}{2009}),
  \bibinfo{pages}{864--870}.
\newblock


\end{thebibliography}

\end{document}